\documentstyle[preprint,prb,aps]{revtex}

\newcommand{\nl}{\nonumber \\}

\newcommand{\be}{\begin{equation}}
\newcommand{\ee}{\end{equation}}
\newcommand{\bea}{\begin{eqnarray}}
\newcommand{\eea}{\end{eqnarray}}

\newcommand{\Eq}[1]{Eq.\,(\ref{#1})}
\newcommand{\Eqs}[1]{Eqs.\,(\ref{#1})}

\newcommand{\ti}{\tilde}

\begin{document}
\draft

\title{Tunneling in partial coherence through a series of barriers:
        an explicit result from a pure dephasing model }

\author{Xin-Qi Li and YiJing Yan}

\address{Department of Chemistry, Hong Kong University of Science and 
         Technology, Kowloon\\ Hong Kong }

\date{\today}

\maketitle
\begin{abstract}
Within the B\"{u}ttiker dephasing model,
the backscattering in the dephasing process
is eliminated by setting a proper boundary condition.
Explicit expression is carried out for the effective total
tunneling probability in the presence of multiple pure dephasing scatterers 
with partial coherence.
The derived formula is illustrated analytically by various 
limiting cases, and numerically for its application in 
tunneling through multi-barrier systems.
\end{abstract}

\vspace{4ex}

\pacs{PACS numbers:73.23.-b,73.40.-c}


To simulate the phase-breaking effect 
in partially coherent transport through a mesoscopic system,
B\"uttiker proposed a conceptually simple model
by coupling electronic reservoirs to the conductor.\cite{But863020,But8863}
The dephasing reservoir can be thought of as either a fictitious 
or a real branch voltage probe.
Although this approach appears to be purely phenomenological,
it however can be justified from a microscopic theory with proper
approximations,\cite{McL9113846,Her9111586,Dat929493} 
by viewing that both the electron-phonon interactions and 
the dephasing reservoir can be described by a self-energy function.
Owing to the simplicity, the B\"uttiker dephasing model
has received noticeable attention,\cite{Dat95,Fer97}
and been applied to transport through various mesoscopic systems.
\cite{DAm907411,Sch92639,Mas942295,Gag9613885,And96270,Mor0067,Liu007645,%
Bee921889,Chr971946,Tex007454,Bla001}

Noticeably, in the original work of B\"uttiker \cite{But863020,But8863}
and the later applications mentioned above, 
in addition to randomizing the electronic phase,
the phase-breaking scatterer would also 
randomize the electronic {\it momentum}.
Randomization of momentum means backscattering in the dephasing process,
thus introduces an additional resistance.
This undesired feature has been noticed and analyzed by a few authors,
\cite{But91213,But9535,Dat95a,Kni99916} commonly 
following the idea by coupling {\it two voltage probes} to model a single 
{\it pure dephasing} scatterer.
In this paper, based on the original work of B\"uttiker 
(i.e.\ using a {\it single reservoir} to model
a single dephasing scatterer),\cite{But863020,But8863}
an explicit expression will be derived for the effective total
tunneling probability through a multi-barrier mesoscopic system
in the presence of multiple pure dephasing scatterers with arbitrary
dephasing strength. 
The underlying physics and practical application will be illustrated
clearly.


In general, consider the tunneling through a series of barriers shown 
in Fig.\ 1, where the squares stand for tunnel barriers,
and the triangles for dephasing scatterers.
They can be described in terms of scattering matrices as follows.
For the individual (symmetric) barrier (e.g.\ the $j$th one), 
the tunneling property is characterized by \cite{Mer98}
\be\label{Selastic}
\underline{\bf S}_j^{\rm el} = \left[ \begin{array}{cc}
	     r_j & t_j     \\ t_j & r_j  \end{array} \right]\, ,
\ee
which connects the two incoming amplitudes with the two outgoing amplitudes
as is well known. 
$t_j$ and $r_j$ are respectively the transmission and reflection coefficients
through the barrier, and can be characterized by a real parameter 
($0\leq\delta_j\leq 1$)
as $t_j=\sqrt{\delta_j}$, and $r_j=i \sqrt{1-\delta_j}$. 
For each dephasing scatterer, 
following B\"uttiker's approach \cite{But863020,But8863},
a dephasing reservoir is coupled to the system via two channels.
During tunneling process, the electron has certain quantum probability of
being scattered into the reservoir, undergoing phase randomization in it,
then returning into the system via the coupler.
As a result of dephasing, 
the re-emitted component does not interfere with 
that having not entered the reservoir.
As a specific model, the coupler is described by the following 
scattering matrix \cite{But863020,But8863}
\be \label{Sepsilon}
\underline{\bf S}_{\epsilon} = \left[ \begin{array}{cccc}
	     0 & \sqrt{1-\epsilon} & \sqrt{\epsilon} & 0  \\
           \sqrt{1-\epsilon} & 0 & 0 & \sqrt{\epsilon}   \\
           \sqrt{\epsilon} & 0 & 0 & -\sqrt{1-\epsilon}  \\
           0 & \sqrt{\epsilon} & - \sqrt{1-\epsilon} & 0 
           \end{array} \right]\, .
\ee
This $4\times 4$ matrix connects the $4$ incoming amplitudes
with the $4$ outgoing amplitudes along the attached 4 channels.
Here, $\epsilon\in [0,1]$ is a parameter characterizing the 
dephasing strength ranging from complete coherence to complete incoherence.

To carry out the effective tunneling probability 
from channel 1 to channel 2 shown in Fig.\ 1, 
the $(2N+2)\times(2N+2)$ S-matrix $\underline{\bf S}^{(N)}$ 
of the entire system is needed.
Note that the dimension of $\underline{\bf S}^{(N)}$
corresponds to the total number of external channels.
By applying either a direct matrix algebra or
the Feynman path technique,
$\underline{\bf S}^{(N)}$ can be carried out via
the following recursive expressions:\cite{Dat95,Sch92639}
\begin{mathletters}\label{SN}
\bea
S^{(N)}_{iI} &=& \ti{s}^{(N)}_{i1} \left[Z_N\chi\right] S^{(N-1)}_{2I} \, , \\
S^{(N)}_{Ii} &=& S^{(N-1)}_{I2} \left[Z_N\chi\right] \ti{s}^{(N)}_{1i} \, , \\
S^{(N)}_{IJ} &=& S^{(N-1)}_{IJ}+S^{(N-1)}_{I2}
	  \left[Z_N\chi\ti{s}^{(N)}_{11}\chi\right] S^{(N-1)}_{2J}  \, ,   \\
S^{(N)}_{ij} &=& \ti{s}^{(N)}_{ij}+\ti{s}^{(N)}_{i1}
	  \left[Z_N\chi S^{(N-1)}_{22}\chi \right]\ti{s}^{(N)}_{1j}   \,  .
\eea
\end{mathletters}
Here  $I,J=1,3,4,\cdots,2N-1,2N$, and $i,j=2,2N+1,2N+2$. 
$\underline{\bf S}^{(N-1)}$ is the scattering matrix 
for the subsystem of $N$ tunnel barriers 
(correspondingly, $N-1$ dephasing scatterers);
and the $4\times 4$ scattering matrix $\ti{\underline{\bf s}}^{(N)}$ is 
for the $N$th single scatterer--barrier segment.
$Z_N=[1-\chi S^{(N-1)}_{22}\chi \ti{s}^{(N)}_{11}]^{-1}$, resulting
physically from the multiple reflections.\cite{Dat95}
$\chi$ denotes the plane-wave propagation-induced phase-change factor 
for the electron moving between the two composite subsystems.
For simplicity, we hereafter set $\chi=1$, which is valid for 
far off-resonance tunneling.

In practice of computing the S-matrix of the entire system, 
we start with $\underline{\bf S}^{(1)}$
for the simple system of barrier-scatterer-barrier. 
Applying the recursive technique described above and making use of \Eqs{Selastic}
and (\ref{Sepsilon}), we obtain
\bea\label{S1}
\underline{\bf S}^{(1)} = Z_1  \left[ \begin{array}{cccc}
r_1+\alpha^2r_2 & \alpha t_1t_2 & \beta t_1 & \alpha\beta t_1r_2    \\
\alpha t_1t_2 & r_2+\alpha^2r_1 & \alpha\beta r_1t_2 & \beta t_2  \\
\beta t_1 & \alpha\beta r_1t_2 & \beta^2r_1 & -\alpha(1-r_1r_2)  \\
\alpha\beta t_1r_2 & \beta t_2 & -\alpha(1-r_1r_2) & \beta^2r_2 	      
                               \end{array} \right] . 
\eea
Here $\alpha=\sqrt{1-\epsilon}$ and $\beta=\sqrt{\epsilon}$ 
are introduced to simplify the notation, 
and $Z_1=(1-\alpha^2r_1r_2)^{-1}$.
The most important advantage of the present technique is its ability
in treating multiple dephasing scatterers.
To this end, we need the S-matrix of the added segment of scatterer-barrier.
It can be easily obtained from \Eq{S1} by setting $r_1=0$ and $t_1=1$,
and replacing $t_2$ ($r_2$) by $t_j$ ($r_j$) for the $j$th coherent barrier.
We denote the resulting 
$4\times 4$ S-matrix by $\ti{\underline{\bf s}}^{(j)}$ [c.f. \Eq{SN}].
Using \Eq{S1} and the recursive rules of
\Eq{SN}, we can therefore evaluate the S-matrix $\underline{\bf S}^{(N)}$ for
the entire system, thus calculate the effective tunneling probability
through the system (i.e.\ from channels 1 to 2).
In the following, for the convenience of comparison, 
we first present result under the {\it momentum randomization}
boundary condition. 
Then, the formal result and its implication/application
based on the {\it pure dephasing} boundary condition
are carried out comparatively.



{\it Momentum Randomization Boundary Condition}.
Under this boundary condition, 
the re-emitting current from the $n$th reservoir into the system 
is assumed to be equally injected 
through its $(2n+1)$th and $(2n+2)$th channels.
We denote this current as $\ti{J}_n$.
In the dephasing process, the electron number is conserved, thus the net 
current of the two channels connecting the system and the reservoir is zero.
This feature is characterized by 
\bea \label{JB1}
(2\underline{\bf I}-\ti{\underline{\bf T}})
  \ti{\underline{\bf J}}_B = \ti{\underline{\bf K}}^{(1)}J_L \, ,
\eea
where $\underline{\bf I}$ is the $N\times N$ unit matrix,
and $\ti{\underline{\bf J}}_B=(\ti{J}_1,\ti{J}_2,\cdots,\ti{J}_N)^T$.
Hereafter the superscript $(\cdots)^T$ means transposition of matrix.
$J_L$ is the current injected from the left channel 1.
$\ti{\underline{\bf T}}$ is a $N\times N$ matrix with elements 
$ \ti{T}_{nm}= T_{2n+1,2m+1}+T_{2n+1,2m+2} 
      +T_{2n+2,2m+1}+T_{2n+2,2m+2} $,
where $T_{i,j}\equiv T_{ij}=|S_{ij}|^2$.
$\ti{\underline{\bf K}}^{(1)}$ is a $N\times 1$ matrix with elements
$ \ti{K}^{(1)}_{n}= T_{2n+1,1}+T_{2n+2,1} $.
Substituting the formal solution of \Eq{JB1},
$\ti{\underline{\bf J}}_B
 = (2\underline{\bf I}-\ti{\underline{\bf T}})^{-1}
   \ti{\underline{\bf K}}^{(1)}J_L $,
into the expression of the outgoing current in the right channel 2,
$J_R=T_{21}J_L+\ti{\underline{\bf K}}^{(2)}\ti{\underline{\bf J}}_B$, 
we obtain the effective tunneling probability under the 
momentum randomization boundary condition
\bea \label{Teff1}
\ti{T}_{\rm eff}\equiv J_R/J_L 
    = T_{21}+ \underline{\ti{\bf K}}^{(2)} 
      (2\underline{\bf I}-\ti{\underline{\bf T}})^{-1}
	      \underline{\ti{\bf K}}^{(1)}  \, ,
\eea
where $\underline{\ti{\bf K}}^{(2)}$
is an $1\times N$ matrix with elements
$\ti{K}_n^{(2)}=T_{2,2n+1}+T_{2,2n+2}$.


{\it Backscattering-Free Boundary Condition}.
To elucidate this boundary condition, 
let us consider in more detail the scattering on the dephasing scatterer
(e.g.\ the $n$th one), see Fig.\ 2.
The scattering matrix $\underline{\bf S}_{\epsilon}$ of \Eq{Sepsilon}
relates the outgoing amplitudes 
$\underline{\bf a}_n'=(a'_{2n+1},a'_{2n+2},b'_{2n+1},b'_{2n+2})^T$
to the incoming amplitudes
$\underline{\bf a}_n=(a_{2n+1},a_{2n+2},b_{2n+1},b_{2n+2})^T$
in terms of 
$\underline{\bf a}_n'=\underline{\bf S}_{\epsilon}\underline{\bf a}_n$.
It is easy to check that
the unitary property of $\underline{\bf S}_{\epsilon}$
and its specific structure would lead to
the conserving relations
\begin{mathletters}\label{pc1}
\bea
|a'_{2n+1}|^2+|b'_{2n+2}|^2 &=& |a_{2n+2}|^2+|b_{2n+1}|^2 \, , \\
|a'_{2n+2}|^2+|b'_{2n+1}|^2 &=& |a_{2n+1}|^2+|b_{2n+2}|^2 \, .
\eea
\end{mathletters}

Now we make use of the {\it pure dephasing} condition:
$|a_{2n+1}|^2=|a'_{2n+2}|^2$, and $|a_{2n+2}|^2=|a'_{2n+1}|^2$.
Its physical meaning is quite clear 
by noting that {\it the phase-breaking only randomizes
the phases of the forward and backward going waves, but 
does not cause any reflections }. Because of this nature, 
we term it as a {\it backscattering-free boundary condition},
which straightforwardly leads to 
\begin{mathletters}\label{pc2}
\bea
|b'_{2n+1}|^2 &=& |b_{2n+2}|^2  \, , \\
|b'_{2n+2}|^2 &=& |b_{2n+1}|^2  \, .
\eea
\end{mathletters}
With this physical insight, we are ready to derive the effective
tunneling probability under the present boundary condition.
We denote $J_n=(j_{2n+1},j_{2n+2})^T$ for the injecting currents from the $n$th
reservoir into the system via the $(2n+1)$th and $(2n+2)$th channels, 
and $J'_n=(j'_{2n+1},j'_{2n+2})^T$ for the currents being scattered into 
the reservoir from the system via the same channels.
The pure dephasing boundary condition, now denoted as 
$j'_{2n+1}=j_{2n+2}$, and $j'_{2n+2}=j_{2n+1}$,
can be expressed compactly for all the $N$ reservoirs as
\bea\label{JB}
{\bf J'}_B=\underline{\bf B}{\bf J}_B ,
\eea
where ${\bf J}_B=(J_1,J_2, \cdots, J_N)^T$, 
${\bf J'}_B=(J'_1,J'_2, \cdots, J'_N)^T$, and 
$\underline{\bf B}=\underline{\bf I}\otimes\underline{\bf \sigma}$.
$\underline{\bf \sigma}$ is a $2\times 2$ matrix with elements
$\sigma_{ij}=1-\delta_{ij}$.
Again, a simple current counting leads to:
\begin{mathletters}\label{JLBR}
\bea
J_R &=& T_{21}J_L +\underline{\bf K}^{(2)} {\bf J}_B \, ,   \\
{\bf J'}_B &=& \underline{\bf K}^{(1)} J_L + \underline{\bf T} {\bf J}_B \, . 
\eea
\end{mathletters}
Here, $\underline{\bf K}^{(1)}$ is a $2N\times 1$ matrix
with elements $T_{n1}$,
and $\underline{\bf K}^{(2)}$ an $1\times 2N$ matrix
with elements $T_{2n}$, with $n=3,4,\cdots,2N+2$.
$\underline{\bf  T}$ is a $2N\times 2N$ matrix with elements
$T_{mn}$, with $m,n=3,4,\cdots,2N+2$.
Straightforwardly, by substituting \Eq{JB} into \Eq{JLBR}
an elegant expression is obtained for the effective
tunneling probability 
\bea \label{Teff2}
T_{\rm eff}\equiv J_R/J_L
  = T_{21} + \underline{\bf K}^{(2)} 
              (\underline{\bf B}-\underline{\bf T})^{-1}
              \underline{\bf K}^{(1)}   \, .
\eea
This equation is formally similar to \Eq{Teff1}. 
Both \Eqs{Teff1} and (\ref{Teff2}) contain a common 
coherent term (the first one), and an incoherent term (the second one).
However, they physically differentiate from each other.
Below we detail our discussions on their difference by 
focusing on the special case of one dephasing scatterer.


In the case of $N=1$, from \Eq{Teff1} we easily obtain 
\bea \label{Teff3}
\ti{T}_{\rm eff}
  &=& T_{21}+(T_{23}+T_{24})(2-\ti{T}_{11})^{-1}(T_{13}+T_{14})  \nl
  &=& T_{21}+ S_b S_f/(S_b+S_f) \, .
\eea
Here, following B\"uttiker,\cite{But863020,But8863} 
the incoherent backward and forward scattering
probabilities are introduced: 
$S_b=T_{13}+T_{14}$, and $S_f=T_{23}+T_{24}$.
In deriving \Eq{Teff3}, the symmetry $T_{mn}=T_{nm}$ and
the sum rule $\sum_{m}T_{mn}=1$ have been used. 
On the other hand, for $N=1$, \Eq{Teff2} leads to
\bea \label{Teff4}
T_{\rm eff} &=& T_{21}+
    [(T_{44}T_{23}T_{31}+T_{33}T_{24}T_{41})   \nl
        & &  +(1-T_{34})(T_{23}T_{41}+T_{24}T_{31})] /Z   \, ,
\eea
where $Z=(1-T_{34})^2-T_{33}T_{44}$.

Further, in the completely incoherent regime ($\epsilon=1$), 
$T_{13}=T_1\equiv |t_1|^2$, $T_{24}=T_2\equiv |t_2|^2$, 
$T_{33}=1-T_1$, and $T_{44}=1-T_2$.
Other $T_{mn}$ in \Eqs{Teff3} and (\ref{Teff4}) are zero.
In this regime, \Eq{Teff3} reduces to
\bea\label{Teff5}
\ti{T}_{\rm eff}
  = \left[ \frac{1}{T_1}+\frac{1}{T_2} \right]^{-1} \, ,
\eea
whereas \Eq{Teff4} gives rise to 
\bea\label{Teff6}
T_{\rm eff}
  = \left[ \frac{1}{T_1}+\frac{1}{T_2}-1 \right]^{-1} \, .
\eea
The latter equation (\ref{Teff6}) can also be obtained 
by a rather simple treatment based on {\it incoherent} multiple reflections 
and transmissions through two barriers.\cite{Dat95b}
The interesting difference between \Eqs{Teff5} and (\ref{Teff6})
is highlighted as follows. 
It is well known that the two-terminal resistance is related to the 
effective total transmission probability 
$T_{\rm tot}$ (i.e.\ $\ti{T}_{\rm eff}$ and $T_{\rm eff}$)
via the Laudauer formula,\cite{Lan57223}
${\cal R}_{\rm tot}=(h/e^2)(1/T_{\rm tot})$.
From \Eq{Teff5} the system resistance can be expressed as
\bea \label{Rtot1}
\ti{\cal R}_{\rm tot} = {\cal R}_1+{\cal R}_2+{\cal R}_c+{\cal R}_s \,.
\eea
In this decomposed form, ${\cal R}_j=(h/e^2)(1-T_j)/T_j$ ($j=1,2$) is the 
Landauer resistance for the $j$th conductor, 
${\cal R}_c=h/e^2$ is the so-called contact resistance rooted in 
the two-terminal configuration (measurement).\cite{Dat95,Imr86101}
Interestingly, the dephasing scatterer contributes a
{\it constant resistance} ${\cal R}_s=h/e^2$
in the completely incoherent regime, due to the 
momentum randomization. 
On the other hand, from \Eq{Teff6} we have 
\bea \label{Rtot2}
{\cal R}_{\rm tot} = {\cal R}_1+{\cal R}_2+{\cal R}_c \,.
\eea
We see that under the boundary condition \Eq{JB}
the dephasing scatterer only plays a phase-breaking role.
As a result, the {\it Landauer resistances} are connected 
in series in a purely classical way,
and the dephasing source does not cause additional resistance.

Interestingly, the backscattering-free nature on the dephasing scatterer
can be further elucidated by considering the simple 
transmission through only a dephasing scatterer
(i.e.\ with no tunnel barriers), with arbitrary dephasing strength.
In this case, the relevant transmission coefficients are:
$T_{21}=T_{34}=1-\epsilon$, $T_{13}=T_{24}=\epsilon$,
$T_{33}=T_{44}=T_{14}=T_{23}=0$.
Accordingly, \Eq{Teff3} gives rise to 
$\ti{T}_{\rm eff}=1-\epsilon/2$, whereas \Eq{Teff4} leads to
$T_{\rm eff}=1$.
This result clearly shows the distinct nature of the two
dephasing models: one causes backscattering, another is
backscattering free.

For two resistors connected in partial coherence ($0<\epsilon < 1$),
the simple sum rule of the individual resistances as \Eqs{Rtot1}
and (\ref{Rtot2}) breaks down. 
However, the resistance difference 
$\Delta{\cal R}= (h/e^2)(1/\ti{T}_{\rm eff}-1/T_{\rm eff})$
based on \Eqs{Teff3} and (\ref{Teff4})
is a proper quantity to characterize the additional resistance 
caused by the dephasing scatterer
under the the momentum randomization boundary condition.
Figure 3 shows $\Delta{\cal R}$ as a function of the dephasing strength
$\epsilon$.
In general, the interplay between the backscattering 
on the dephasing scatterer and the tunneling through the individual barriers
leads to $\Delta{\cal R}(\epsilon)$ depending on the 
barrier-tunneling strength as shown in Fig.\ 3.

To further illustrate the application of \Eqs{Teff1} and (\ref{Teff2})
in combination with the recursive rule of \Eq{SN},
we briefly present results for tunneling in the presence of 
multiple phase-breaking scatterers.
For clarity, Figure 4 shows the relative tunneling probabilities versus
the dephasing strength, i.e., $T_{\rm eff}(\epsilon)/T_{\rm eff}(\epsilon=0)$
and $\ti{T}_{\rm eff}(\epsilon)/\ti{T}_{\rm eff}(\epsilon=0)$,
by solid and dashed curves, respectively.
In the weak tunneling regime ($\delta=0.1$) shown in Fig.\ 4(a),
dephasing enhances the tunneling remarkably, and the 
two dephasing models, i.e., \Eqs{Teff1} and (\ref{Teff2}),
give almost the same results.
However, in the strong tunneling regime ($\delta=0.9$) shown in Fig.\ 4(b),
the two dephasing models give noticeably different results.
More interestingly, the turnover behavior predicted by \Eq{Teff1},
which was paid special attention by B\"uttiker,
\cite{But863020} does not occur in the pure dephasing model of \Eq{Teff2}.
Therefore, the turnover behavior appears a consequence of competition
between the incoherent-tunneling and the dephasing-induced backscattering.
It disappears after the backscattering being eliminated.

In summary, we have presented a unified treatment for  
phase-breaking tunneling based on the B\"ttiker model.
Simple adopting of two types of boundary conditions
can clear up the distinct natures of two dephasing models.
The derived explicit expressions in combination with the 
recursive rules can be conveniently applied to tunneling
through multi-barrier systems in arbitrary partial coherence.

\section*{Acknowledgments}
    Support from the Research Grants Council of the Hong Kong Government
and the National Natural Science Foundation of China is gratefully
acknowledged.



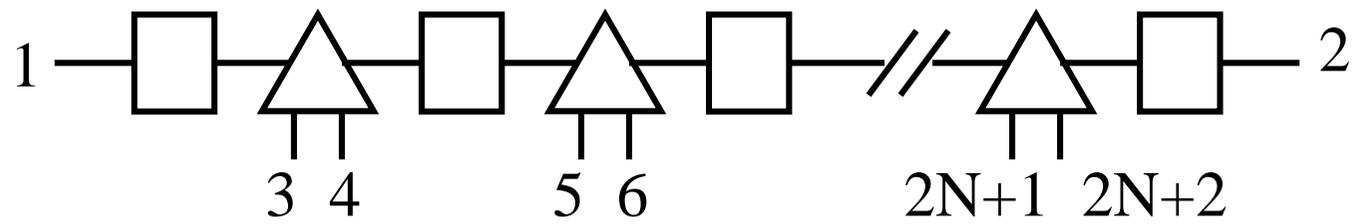
\begin{figure}\label{Fig1}
\caption{ Tunneling in partial coherence through a series of barriers.
The square represents a tunnel barrier, and the triangle stands for
a phase-breaking scatterer which couples the system to a dephasing
reservoir via two channels. }
\end{figure}

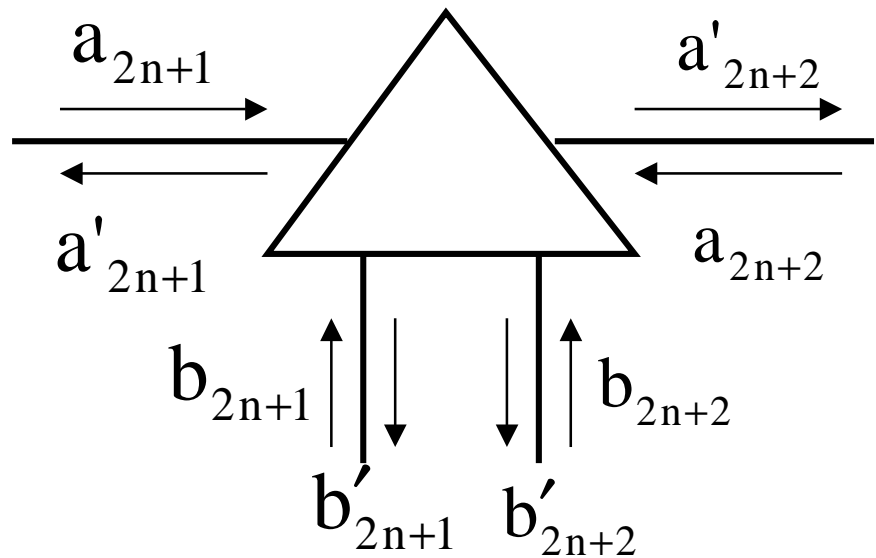
\begin{figure}\label{Fig2}
\caption{ Schematic diagram for the scattering on the $n$th dephasing
scatterer. The outgoing amplitudes are connected with the
incoming amplitudes via 
$\underline{\bf a}_n'=\underline{\bf S}_{\epsilon}\underline{\bf a}_n$,
see text for more details.  }
\end{figure}

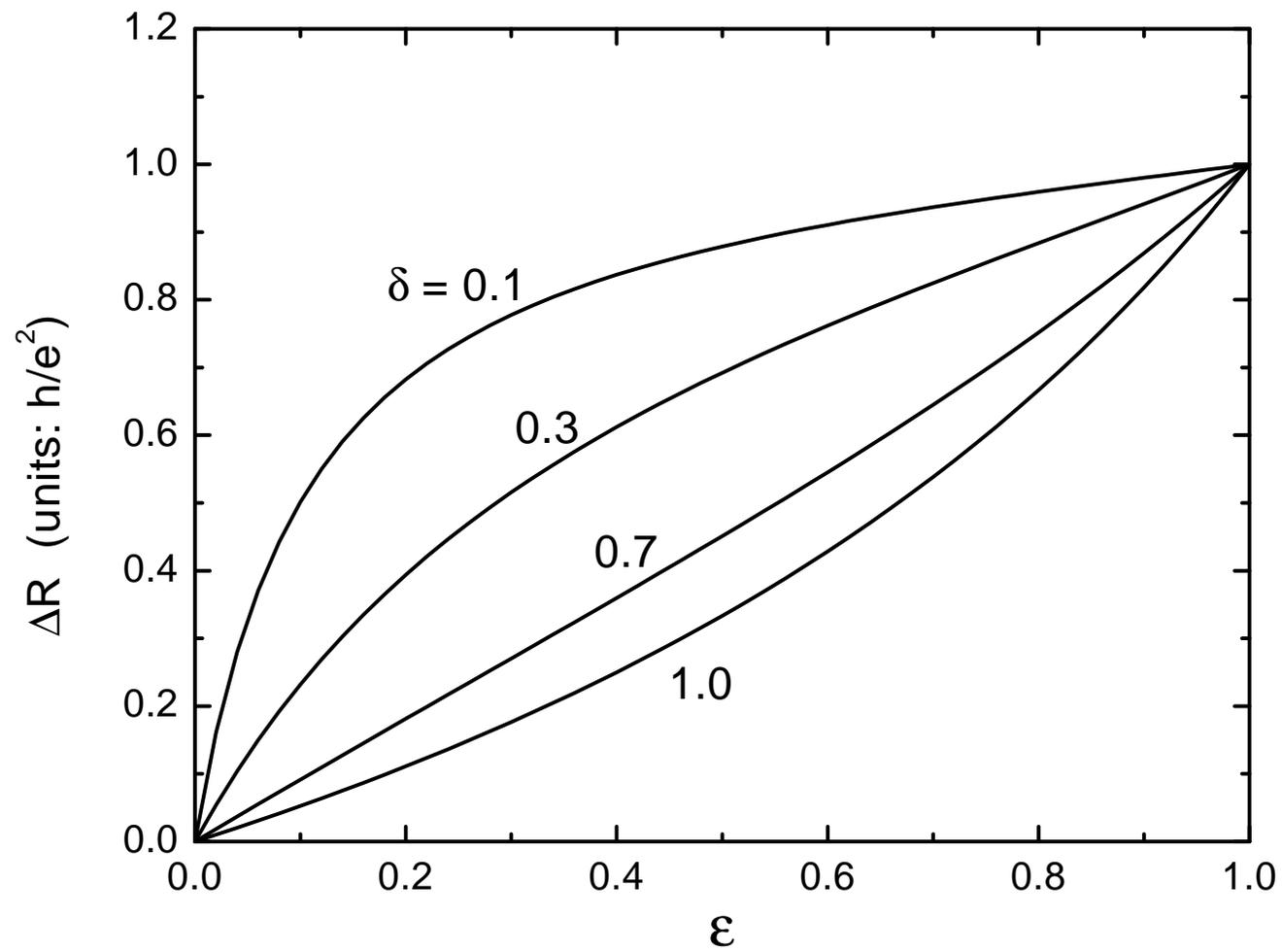
\begin{figure}\label{Fig3}
\caption{ Additional resistance caused by momentum randomization 
on the dephasing scatterer. Here the special case of 
two identical resistors connected in partial coherence is demonstrated
to show the interplay between dephasing and tunneling.
$\delta$ characterizes the tunneling strength through
the individual resistor, i.e.\ $T_1=T_2=\delta$.  }
\end{figure}

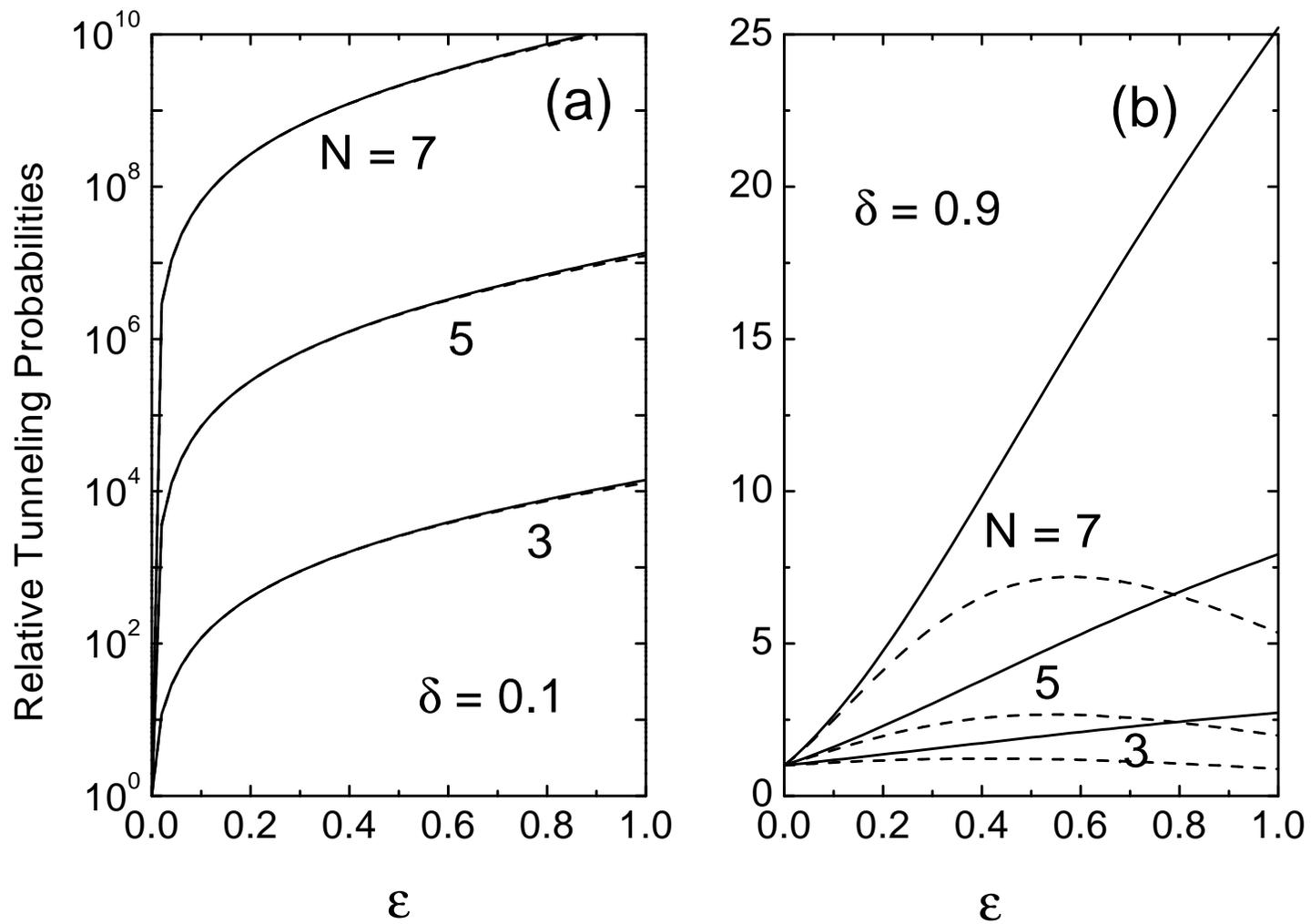
\begin{figure}\label{Fig4}
\caption{ Partial-coherence tunneling through $N+1$ identical barriers
(i.e.\ $T_j=\delta$, $j=1,2,\cdots,N+1$). 
The relative tunneling probabilities 
$T_{\rm eff}(\epsilon)/T_{\rm eff}(\epsilon=0)$
and $\ti{T}_{\rm eff}(\epsilon)/\ti{T}_{\rm eff}(\epsilon=0)$
are respectively shown by the solid and dashed curves.
The weak (a) and strong (b) tunneling limits are plotted
to highlight the similarity and difference of the two 
dephasing models in different regimes.   }
\end{figure}


\begin{references}
\bibitem{But863020}
M.~B\"uttiker, Phys. Rev. B {\bf 33}, 3020 (1986).
\bibitem{But8863}
M.~B\"uttiker, IBM J. Res. Develop. {\bf 32}, 63 (1988).
\bibitem{McL9113846}
M.~J. McLennan, Y. Lee, and S. Datta,
Phys. Rev. B {\bf 43}, 13846 (1991).
\bibitem{Her9111586}
S. Hershfield, Phys. Rev. B {\bf 43}, 11586 (1991).
\bibitem{Dat929493}
S. Datta, Phys. Rev. B {\bf 46}, 9493 (1992).

\bibitem{Dat95}
S. Datta, {\it Electronic Transport in Mesoscopic Systems}
(Cambridge University Press, New York, 1995).
\bibitem{Fer97}
D.~K. Ferry and S.~M. Goodnick, {\it Transport in nanostructures} 
(Cambridge University Press, New York, 1997)
\bibitem{Mer98}
E.~Merzbacher, {\it Quantum Mechanics}
(Wiley, New York, 3rd edition, 1998)

\bibitem{DAm907411}
J.L. D'Amato and H.M. Pastawski, 
Phys. Rev. B {\bf 41}, 7411 (1990).
\bibitem{Sch92639}
M. Schreiber and M. Maschke, Philos. Mag. {\bf 65}, 639 (1992).
\bibitem{Mas942295} 
K. Maschke and M. Schreiber, Phys. Rev. B {\bf 49}, 2295 (1994).
\bibitem{Gag9613885}
F. Gagel and K. Maschke, Phys. Rev. B {\bf 54}, 13885 (1996).
\bibitem{And96270}
T. Ando, Surface Science {\bf 361/362}, 270 (1996).
\bibitem{Mor0067}
N.A. Mortensen, A-P. Jauho, and K. Flensberg,
Superlatt. Microstruct. {\bf 28}, 67 (2000).
\bibitem{Liu007645}
M.T. Liu and C.S. Chu, Phys. Rev. B {\bf 61}, 7645 (2000).
\bibitem{Bee921889}
C.W.J. Beenakker and M. B\"uttiker,
Phys. Rev. B {\bf 46}, 1889 (1992).
\bibitem{Chr971946}
T. Christen and M. B\"uttiker, Phys. Rev. B {\bf 55}, R1946 (1997).
\bibitem{Tex007454}
C. Texier and M. B\"uttiker, Phys. Rev. B {\bf 62}, 7454 (2000).
\bibitem{Bla001}
Ya.M. Blanter and M. B\"uttiker, Phys. Rep. {\bf 336}, 1 (2000).


\bibitem{But91213}
M. B\"uttiker, in {\it Resonant Tunneling in Semiconductors: 
Physics and Applications}, edited by L.L. Chang, E.E. Mendez, 
and C. Tejedor (Plenum, New York, 1991), p.\ 213.
\bibitem{But9535}
M. B\"uttiker, in {\it Proceedings of the 13th International Conference
on Noise in Physical Systems and 1/f Fluctuations},
edited by V. Bareikis and R. Katilius 
(World Scientific, Singapore, 1995), p.\ 35.
\bibitem{Dat95a}
See Ref.\ 6, pp.\ 129-132.
\bibitem{Kni99916}
I. Knittle, F. Gagel, and M. Schreiber,
Phys. Rev. B {\bf 60}, 916 (1999).

\bibitem{Dat95b}
See Ref.\ 6, p.\ 64.

\bibitem{Lan57223}
R.~Landauer, IBM J. Res. Dev. {\bf 1}, 223 (1957);
Philos. Mag. {\bf 21}, 863 (1970).
\bibitem{Imr86101}
Y.~Imry, \newblock ``Physics of mesoscopic systems,'' in {\em Directions in
  Condensed Matter Physics: Memorial Volume in Honor of Prof.~S.~K.~Ma}, edited
  by G.~Grinstein and G.~Mazenko, pages 101--163, World Scientific, Singapore,
  1986.

\end{references}
\end{document}